\documentclass[%
 aip,
 amsmath,amssymb,
 reprint,
]{revtex4-1}

\usepackage{graphicx}
\usepackage{dcolumn}
\usepackage{bm}

\usepackage[utf8]{inputenc}
\usepackage[T1]{fontenc}
\usepackage{mathptmx}

\begin{document}

\preprint{AIP/123-QED}

\title[High finesse microcavities in the optical telecom O-band]{High finesse microcavities in the optical telecom O-band}

\author{J. Fait}
 \altaffiliation{These authors contributed equally to this work}
\affiliation{ 
Vienna Center for Quantum Science and Technology, Universität Wien, Boltzmanngasse 5, 1090 Vienna, Austria
}%
\affiliation{ 
Institute of Physics, Academy of Sciences of the Czech Republic, Cukrovarnická 10, 162 00 Prague, Czech Republic
}%
\affiliation{ 
Czech Technical University in Prague, Faculty of Electrical Engineering, Technická 2, 166 27 Prague, Czech Republic
}%

\author{S. Putz}%
 \altaffiliation{These authors contributed equally to this work}
\affiliation{ 
Vienna Center for Quantum Science and Technology, Universität Wien, Boltzmanngasse 5, 1090 Vienna, Austria
}%

\author{G. Wachter}%
\affiliation{ 
Vienna Center for Quantum Science and Technology, Universität Wien, Boltzmanngasse 5, 1090 Vienna, Austria
}%

\author{J. Schalko}%
\affiliation{ 
Institute of Sensor and Actuator Systems, TU Wien, Gußhau\ss str. 27-29, 1040 Vienna, Austria
}%

\author{U. Schmid}%
\affiliation{ 
Institute of Sensor and Actuator Systems, TU Wien, Gußhau\ss str. 27-29, 1040 Vienna, Austria
}%

\author{M. Arndt}%
\affiliation{ 
Vienna Center for Quantum Science and Technology, Universität Wien, Boltzmanngasse 5, 1090 Vienna, Austria
}%

\author{M. Trupke}
 \altaffiliation{Correspondence should be sent to: michael.trupke@univie.ac.at}
\affiliation{ 
Vienna Center for Quantum Science and Technology, Universität Wien, Boltzmanngasse 5, 1090 Vienna, Austria
}%

\date{\today}

\begin{abstract}
Optical microcavities allow to strongly confine light in small mode volumes and with long  photon lifetimes. This confinement significantly enhances the interaction between light and matter inside the cavity, with applications such as optical trapping and cooling of nanoparticles, single-photon emission enhancement, quantum information processing, and sensing. For many applications, open resonators with direct access to the mode volume are necessary. Here we report on a scalable, open-access optical microcavity platform with mode volumes \(< 30 \lambda^3\) and finesse approaching \( 5\times10^5\). This result significantly exceeds the highest optical enhancement factors achieved to date for Fabry-Pérot cavities. The platform provides a building block for high-performance quantum devices relying on strong light-matter interaction.
\end{abstract}

\maketitle

\section{\label{sec1}Introduction}

The confinement of the electromagnetic field inside a small volume is key in building devices with high quantum efficiency, by enhancing the interaction strength between photons and matter. For instance, cavities can increase the single-photon count rate from an emitter due to the Purcell effect\cite{purcell1946}. More generally, the strong confinement enables coherent interactions of light and atoms in cavity quantum electrodynamics (CQED)\cite{thompson1992,Reiserer2015}, or with dielectric particles in cavity quantum optomechanics\cite{delic2020}.

In many applications, the performance of the device depends on the spatial confinement of the electromagnetic field given by the volume of the cavity mode \(V\) and the temporal confinement of the field given by the quality factor \(Q\). Thus the figure of merit is the optical enhancement given by the relation
\begin{eqnarray}
\Upsilon=\frac{Q}{V}
\left(\frac{\lambda}{n}\right)^{3}
\label{eq:one}
\end{eqnarray}
with the refractive index of the cavity medium \(n\) and the wavelength \(\lambda\). This term is a key quantity in the coherent interaction of nanoparticles, atoms, or molecules with photons. An important example is the well known Purcell factor, which can be expressed in terms of the optical enhancement factor as \(P=3\Upsilon\eta/4\pi^2\), where \(\eta\) is the branching ratio of the relevant two-level transition.

\begin{figure*}[t!]
\includegraphics{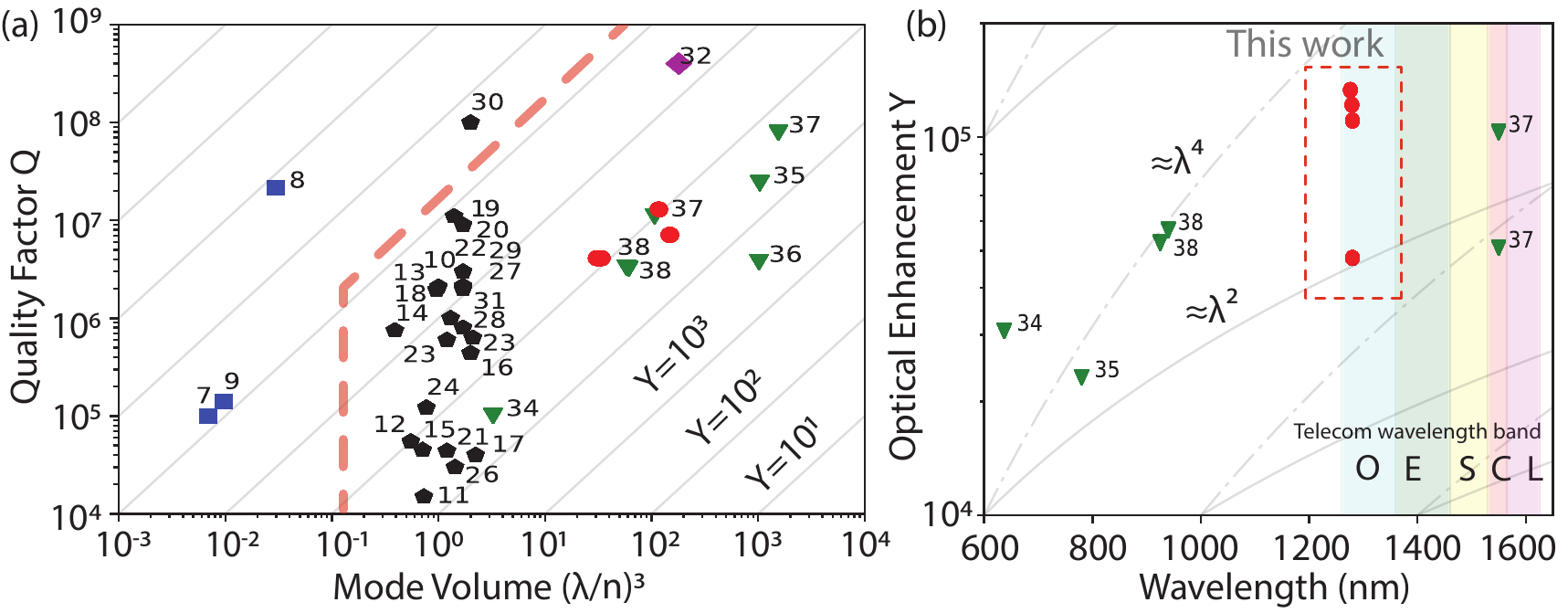}
\caption{\label{fig1-comparison}(a) Dependence of the cavity quality factor \(Q\) on the cavity mode volume \(V\) for microcavities with optical enhancement \(\Upsilon > 10^4\): bowtie and slotted photonic cavities – blue squares,\cite{hu2018, sun2017, seidler2013} conventional PhC cavities – black pentagons\cite{maeno2017, balet2007, khan2011, lai2014, deotare2009, akahane2003, zhan2020, song2018, han2010, asano2017, sekoguchi2014, hostein2009, taguchi2011, song2005, debnath2017, song2019, gonzalez2013, terawaki2012, kuramochi2006, kuwabara2019,Notomi08,Simbula2017}, microtoroid cavity – magenta diamond \cite{kippenberg2004}, FP cavities – green triangles \cite{bogdanovic2017, najer2017, uphoff2015, Colombe2007, wachter2019, muller2010}, and FP cavities presented in this work – red dots (see text).
 The grey lines mark the constant optical enhancement. The red dashed line indicates calculated bounds for FP cavities based on the smallest possible mode volume possible\cite{Kelkar15} and the highest finesse achieved in a FP cavity\cite{Rempe92} assuming that \(Q = {\cal F}\) for \(V_{min} = \lambda^3/8\). (b) The optical enhancement factor of all FP cavities in (a) and their dependence on the wavelength. Gray lines indicate the expected scaling of the optical enhancement factor with the wavelength due to \(\lambda^2\) purely geometric scaling and  \(\lambda^4\) surface roughness and geometric scaling \cite{muller2010}.}
\end{figure*}

Many types of optical resonators exist, such as Fabry-Pérot (FP), micro-sphere, micro-disc or photonic crystal (PhC) cavities, that can be used to enhance light-matter interactions \cite{vahala2003,Janitz2020}. The comparison of \(Q\) and \(V\) values for several types of microcavities found in the literature is shown in Fig.~\ref{fig1-comparison}a.

PhC cavities confine light in a high index material and typically feature a mode volume on the order of \((\lambda/n)^3\). There are numerous realizations of PhC cavities, ranging from 2D PhCs with missing holes, PhC waveguides with local modification of the holes' position or diameter, or 1D PhC nanobeam cavities. The smallest optical mode volumes, far smaller than a cubic wavelength, can be achieved in bowtie photonic crystal structures\cite{hu2018}, which enable a maximal optical enhancement on the order of \(10^8\). The monolithic structure of PhC cavities reduces the mechanical noise and the presence of an optical bandgap may reduce undesired radiative transitions.

The optical performance in photonic crystal cavities is outstanding for many applications, but they also present several practical  limitations. Spectral tuning of the PhC cavity modes is cumbersome and requires technically challenging procedures, e.g. post-fabrication etching \cite{kim2018}. This prevents fast response to changes in the surrounding environment required for frequency locking and stabilization. Approaches to achieve \emph{in-situ} tunable operation have been implemented. However, these methods mostly enable only slow tuning or a moderate tuning range\cite{Petruzzella2018}. Moreover, PhC cavities form closed systems and emitters must be placed directly inside the PhC cavity material to achieve maximal coupling. Precise placement of emitters in PhC cavities is extremely challenging, and the emitters' properties may be negatively affected by material defects and stress originating from the fabrication process\cite{riedrich-moller2014,riedrich-moller2015}. The performance of spin centers can furthermore be adversely impacted in PhC structures by the unavoidable proximity to the surface\cite{Bernardi2018}. The emitters can be placed outside the PhC cavity, albeit at the cost of a reduced coupling strength and additional optical losses\cite{zhu2010}. Finally, coupling of light into and out of the PhC cavity requires careful mode matching and poses significant challenges for achieving a high collection efficiency of light at the output. Similar issues affect micro-toroid and micro-sphere cavities which have larger mode volumes, usually on the order of several tens of \(\lambda^3\), compensated by correspondingly higher Q-factors\cite{kippenberg2004}.

In contrast, FP cavities offer high versatility, enabling an open-access system that can be tuned both broadly and precisely by changing the mirror spacing, which can be done in a scalable manner \cite{derntl2014}. The open-access cavity allows to place atoms, molecules, nanoparticles, or thin solid membranes inside the cavity. However, this comes at the expense of larger cavity mode volumes when compared to PhC cavities, which cannot be compensated fully by the generally higher quality factors of FP cavities (See Fig. ~\ref{fig1-comparison}a). We increase the calculated cavity mode volume of FP cavities by assuming that the mode penetrates to a depth of 0.8\(\lambda\) into each mirror \cite{hood2001} and the cavity is effectively longer than what follows from the free spectral range.

Here, we report on progress in creating a scalable architecture for open-access FP micro-cavities with high quality factors and small mode volumes. To the best of our knowledge, the reported \(Q/(V/\lambda^3) \approx 1.33 \times 10^5\) value exceeds the performance of all open-access optical microcavities to date \cite{birnbaum2005,najer2017,wachter2019}. Our microcavity mirrors are coated for the telecom O-band wavelength range (\(1.26 \:\mathrm{\mu m}- 1.36 \:\mathrm{\mu m}\)), where high-performance, open-access microcavities have not been demonstrated so far (Fig.~\ref{fig1-comparison}b). We have chosen this band since several promising emitters exist in this spectral region, such as vanadium (V) center in silicon carbide (SiC) \cite{spindlberger2019, wolfowicz2020} and the G and T centers in Si\cite{Hollenbach2020,Redjem2020,Bergeron2020}. SiC and Si are stable host environments and offer superb optical properties in wafer-scale substrates. They are therefore prime candidates for implementation in long range quantum communication networks\cite{Atature2018}.

\section{\label{sec2}Results}

An important figure of merit for FP cavities is the finesse \(\cal F\), which is the ratio of the cavity free spectral range \(\Delta\nu_{FSR}\) to the linewidth of the resonance modes \(\Delta\nu_{FWHM}\), and is directly related to the quality factor \(Q\) and to the total round-trip optical losses \(l_{rt}\):
\begin{eqnarray}
{\cal F}
= \frac{\Delta\nu_{FSR}}{\Delta\nu_{FWHM}}
= \frac{\lambda Q}{2L}
= \frac{\pi}{2\mathrm{arcsin}\left(
  \frac{1-\sqrt{1-l_{rt}}}{2\sqrt[4]{1-l_{rt}}}\right)}
\approx \frac{2\pi}{l_{rt}},
\label{eq:two}
\end{eqnarray}
where \(\lambda\) is the cavity mode wavelength and \(L\) the effective optical length of the cavity (approximately equal to the mirror spacing). The total round-trip losses comprise: scattering losses that depend on the surface roughness of the cavity mirrors, clipping losses caused by the lateral size and shape of the mirrors, absorption losses inside the cavity and by the transmission of the input/output mirrors. The finesse \({\cal F}\) is proportional to the number of round trips before a photon leaves the cavity or is lost via dissipation, while the Q-factor is proportional to the average number of optical cycles before a photon is lost from the cavity, and for the shortest possible cavity where \(L=\lambda/2\), the finesse becomes equal to the Q-factor.

In the paraxial approximation, the effective cavity length \(L\) and the radius of curvature \(R\) of the mirrors determine the mode properties of the resonator. The beam waist \(w_0\) defines the mode volume \(V\) given by
\begin{eqnarray}
V = \frac{\pi}{4}w_0^2L, \; \mathrm{with} \; w_0=\sqrt{\frac{\lambda}{\pi\alpha}\sqrt{\alpha RL-L^2}}
\label{eq:three}
\end{eqnarray}
with the wavelength \(\lambda\) and \(\alpha = \{1,2\}\) for plano-concave (PC) and concave-concave (CC) cavities, respectively. With the Rayleigh range \(z_r = \pi w_0^2/\lambda\) of the cavity mode of the Gaussian beam, the radius of curvature (ROC) is then given by \(R = z[1+(z_r/z)^2]\), where \(z\) is the distance measured longitudinally from the beam waist.

\begin{figure}
\includegraphics{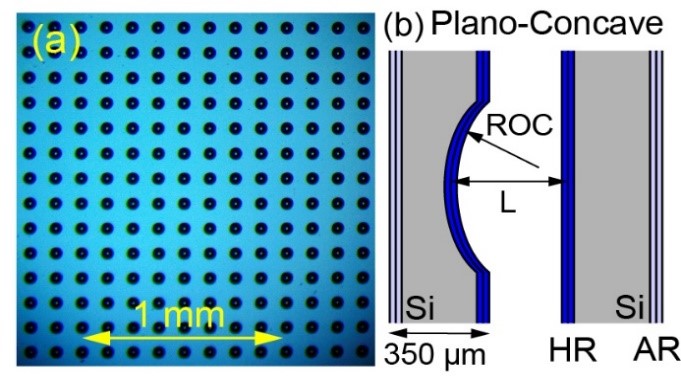}
\caption{\label{fig2-scheme} (a) Microscope image of the coated Si chip showing the micro mirrors on a \(125 \:\mu m\) grid (b) Schematic of a PC-FP microcavity.}
\end{figure}

Short cavities with a small beam waist naturally require a small ROC, which is challenging to fabricate with high precision. Our micromirrors were fabricated from four-inch silicon wafer substrates by a two-step dry etching process, followed by oxidation smoothing (see \citet{wachter2019} for details). The mirror shape was determined using a white light interferometer (Filmetrics Profilm 3D) to extract \(R\). The structured substrate is diced into chips of \(3\times3 \:\mathrm{mm}^2\) which host hundreds of mirrors with a square pitch of \(125\: \mathrm{\mu m}\) (see Fig.~\ref{fig2-scheme}). The Si chips are coated with a dielectric high-reflectivity (HR) Bragg multi-layer coating with a specified transmission of \( 5 \:\mathrm{ppm}\) transmissivity at 1280 nm and an excess coating loss of up to \(1 \:\mathrm{ppm}\) per mirror. The total coating loss between 10 ppm and 12 ppm results in a coating-limited finesse of \(5.2\times 10^5\leq{\cal F_{\text{max}}}\leq 6.3 \times 10^5\). The backside of the Si chips is coated with a broadband anti-reflection (AR) layer. To characterize the cavities, we used a continuously tunable, narrow linewidth laser source (EXFO T100S-HP). The laser light is coupled into the cavities in free space and the reflected laser light is routed through a fiber circulator and detected by a high-bandwidth photodiode.

\begin{figure}
\includegraphics{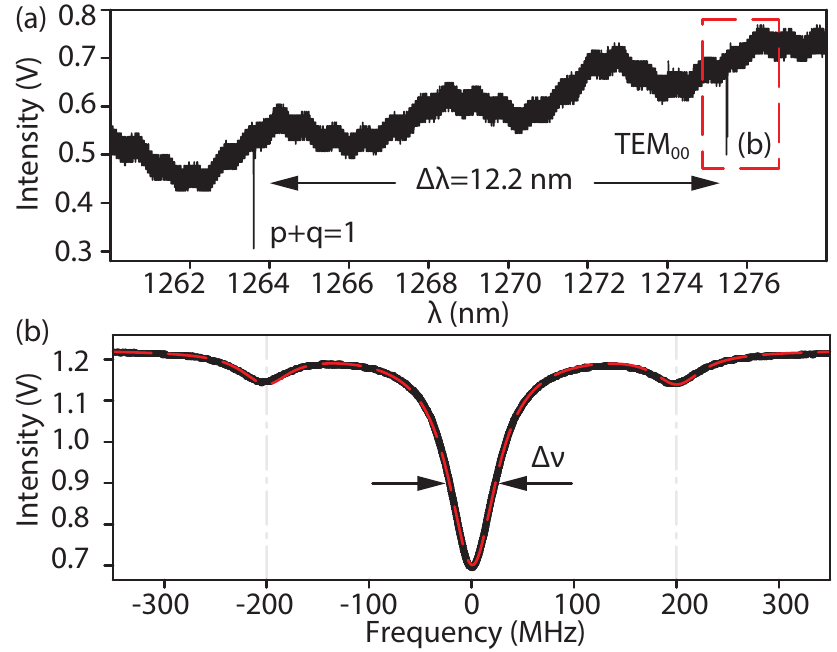}
\caption{\label{fig3-spectra} Spectra for “PC-f“ (see Table~\ref{tab1}): (a) Fundamental \(\mathrm{TEM_{00}}\) mode and one higher order mode. (b) Resonance at \(\lambda=1276\:\mathrm{nm}\) with sideband modulation of 200 MHz (denoted by vertical dash-dotted line). FWHM linewidth of \(\Delta \nu = (58 \pm 2)\:\mathrm{MHz}\) is extracted by a Lorentzian fit (red dashed line).}
\end{figure}

The shortest cavities are assembled in a PC configuration by gluing two mirror chips together. The length of each cavity is fixed, but in order to vary the resonance frequency of the cavities across the array, a spacer with \(\sim 16 \:\mathrm{\mu m}\) thickness was inserted between the Si chips on one side, forming a narrow wedge with \(\sim 0.3\) degrees tilt. The curved mirrors have \(R = (69.3 \pm 8.3)\:\mathrm{\mu m}\) and the depth of the mirrors is  \(\approx 4.5\:\mathrm{\mu m}\). The cavities are characterized by scanning the laser wavelength over a broad range and recording the position and spectral distance between fundamental and higher order modes. The FSR for such short cavities could not be observed directly as it exceeds the wavelength tuning range of the laser. Instead, we determined the effective cavity length from the spectral position of fundamental and higher order modes (see Fig.~\ref{fig3-spectra}a), and from the measured radius of curvature. The ROC is related to the effective cavity length \(L\) by:
\begin{eqnarray}
L = R\left(
  1-\left[
    \mathrm{cos}\left
      (\frac{1}{p+q}\Delta\nu_{p+q}\frac{2L\pi}{c}
    \right)
  \right]^2
\right)
\label{eq:four}
\end{eqnarray}
where \((p,q)\) is the Hermite-Gaussian beam order, \(\Delta\nu_{p+q}\) is the frequency difference between \(\mathrm{TEM}_{0,0}\) and \(\mathrm{TEM}_{p,q}\) Gaussian modes, and \(c\) is the speed of light. With Eq.~(\ref{eq:four}) the calculated effective length of the shortest cavity that supported a \(\mathrm{TEM}_{0,0}\) mode at 1275.7 nm and a mode with \(p+q=1\) at 1263.5 nm was \(L=(6.7 \pm 0.5)\:\mathrm{\mu m}\), which corresponds to \(\Delta\nu_{FSR} = c/2L = (20.3 \pm 1.4)\:\mathrm{THz}\) (see Fig.~\ref{fig3-spectra}a). Hence the mode volume is  \(V= (30.8 \pm 5)\:\lambda^3\). We conservatively reduce the optical enhancement factor for all FP cavities by assuming that the mode penetrates to a depth of \(0.8 \lambda\approx1 \:\mathrm{\mu m}\;\) into each mirror leading to an increase of the mode volume\cite{hood2001}. 

The cavity linewidth was measured by scanning the probe laser over the fundamental \(\mathrm{TEM}_{0,0}\) cavity resonance while a 200 MHz sideband modulation was applied to the carrier, and the observed linewidth is compared to the known splitting of the sidebands (see Fig.~\ref{fig3-spectra}b). The highest observed cavity finesse for the shortest possible cavities was \({\cal F}\approx 3.5\times10^5\) which corresponds to a round-trip loss of 18 ppm. The finesse is \(\sim 45 \% \) lower then the theoretical upper limit and the excess dissipation of \(\approx 8\) ppm is attributed to the combination of slight distortions of the curved mirror shape, the tilt between mirrors, and scattering losses on the mirrors. An overview of all measured cavity configurations can be found in Table~\ref{tab1}.

\begin{table*}[t]
\caption{\label{tab1}Comparison of selected cavity assemblies}
\begin{ruledtabular}
\begin{tabular}{lcccccccc}
Type & \(\lambda\:(\mathrm{nm})\) & \(R\:(\mathrm{\mu m})\) & \(L\:(\mathrm{\mu m})\) &\(\omega_0\:(\mathrm{\mu m})\) & \(V\:(\lambda^3)\) & \({\cal F}\:(/10^3)\) & \(Q\:(/10^6)\)& \(\Upsilon\:(/10^5)\)\\
\hline
PC-f & 1276 & 69.3\(\pm\)8.3 & 8.7\(\pm\)0.7 & 3.05\(\pm\)0.16 & 30.8\(\pm\)5 & 350\(\pm\)30 & 4.1\(\pm\)0.6 & \(\approx\) 1.8\\
PC-f2 & 1279 & 69.3\(\pm\)8.3 & 9.3\(\pm\)0.8 &3.10\(\pm\)0.16& 33.7\(\pm\)5 & 330\(\pm\)20 & 4.1\(\pm\)0.5 & \(\approx\) 1.6\\
PC-a & 1280 & 105.6\(\pm\)17.1 & 18.9\(\pm\)0.1& 4.05\(\pm\)0.18& 116.7\(\pm\)10 & 490\(\pm\)90 & 12.9\(\pm\)2.4 & \(\approx\) 1.3\\
CC-a & 1280 & 105.6\(\pm\)17.1 & 27.4\(\pm\)0.1 & 3.79\(\pm\)0.17& 148.2\(\pm\)14 & 180\(\pm\)10 & 7.1\(\pm\)0.4 & \(\approx\) 0.5\\
\end{tabular}
\end{ruledtabular}
\end{table*}

We also assembled cavities with variable mirror spacing by mounting one of the mirrors on a piezoelectric actuator, enabling rapid tuning of the resonance frequency. Moreover, the chips can be shifted and tilted relative to each other, which allows to assemble CC cavities and also to better compensate for the losses originating from the non-ideal mirror tilt. For this setup we measured PC and CC cavity configurations with  \( L \geq 15 \:\mathrm{\mu m}\). These longer cavities allowed to directly measure the free spectral range and determine the cavity length. We found good agreement between the effective cavity length \(L\) obtained via the FSR and the value computed using Eq.~(\ref{eq:four}). For the actuated cavities we used mirrors with larger \(R=105.6 \:\mathrm{\mu m}\) with a mirror depth of \(8.5\:\mathrm{\mu m}\), giving the minimum spacing between the mirror chips of \(\sim6\:\mathrm{\mu m}\). This configuration reaches \({\cal F}\approx 4.9 \times 10^5\) which is close to \(80 \% \) of the upper theoretical limit (see Table 1: PC-a). The corresponding losses \(13 \pm 3 \:\mathrm{ppm}\) are dominated by the mirror transmissivity. For the CC cavity configuration we observed slightly lower finesse values of \({\cal F}\approx 1.33 \times 10^5\). We presume that this reduction is due to transverse misalignment of the cavity chips with respect to each other.

\begin{figure}
\includegraphics{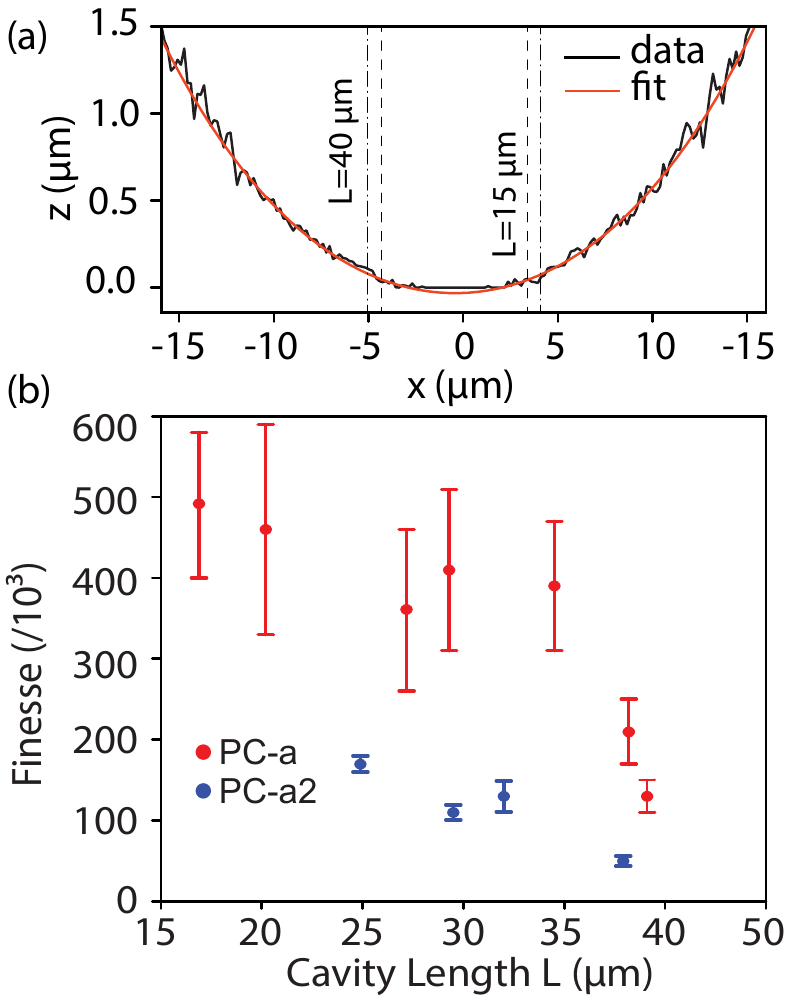}
\caption{\label{fig4-length} (a) White light interferometer measurement used to determine the mirror radius of curvature of mirror PC-a. The fit is a two dimensional parabolic fit with a higher order correction term. The dependence of the finesse on the cavity length for PC-a (red) and PC-a2 (blue, mirror with additional excess losses).}
\end{figure}

The finesse of all measured cavity configurations degrades with increasing cavity length (Table~\ref{tab1}). To measure this systematically we used piezo actuated cavities and measured the optical finesse as a function of cavity the length, Fig.~\ref{fig4-length} shows a similar degradation with increasing cavity length. For cavity lengths between \(\approx 15 - 35 \:\mathrm{\mu m}\), the finesse decreases slowly with increasing cavity length. This behavior can be explained by an increasing beam waist on the curved mirror for longer cavities, which increases the scattering, absorption and clipping losses. This moderate downward trend is followed by a sharp drop of the finesse for cavity lengths longer than \(\approx 35 \:\mathrm{\mu m}\) and no resonance could be observed for cavities longer than \(\approx 40 \:\mathrm{\mu m}\). The losses at \(L\approx 39 \:\mathrm{\mu m}\) amount to 50 ppm and exceed the other types of losses, even considering the imperfections listed above (blue data set in Fig.~\ref{fig4-length}.) The finesse could not be measured for cavities longer than \(\approx 40 \:\mathrm{\mu m}\) as the contrast of modes measured in reflection was too small. It is likely that the extra losses are induced by the deviation of the cavity mirrors from an ideal parabolic shape, which becomes non-negligible for the larger spot sizes resulting from larger mirror separations.

\section{\label{sec3}Conclusion}

In conclusion, we have built high finesse optical Fabry-Perot microcavities in the telecom wavelength O-band at 1280 nm.
We showed that the finesse remains high for a cavity length of up to \(\approx 35 \:\mathrm{\mu m}\) or \(L/R \approx  0.35\), where losses due to a non the parabolic mirror shape become dominant. The cavities are suitable for enhancement of photon count rates from vanadium in silicon carbide (SiC) or G centers in silicon. The extracted optical enhancement value reaches \(\Upsilon=1.33\times10^5\), more than two times larger than other types of open-access microcavities. This improvement is highly desirable for spin-photon interfaces, cavity cooling of nanoparticles, and other applications requiring extreme enhancement of the interaction of light and matter.

\section{\label{sec4}Acknowledgement}
J.F. acknowledges funding by the Czech Science Foundation GAČR (19-14523S) and OeAD-GmbH (Aktion Österreich-Tschechien program, ICM-2019-13725); S.P. acknowledges support by the Marie Skłodowska‐Curie Action through the Erwin Schr{\"o}dinger Quantum Fellowship Program (No 801110). M. T. acknowledges FWF projects I 3167-N27 and EU FET-Open 862721 QuanTELCO.

\section{\label{sec5}References}

\nocite{*}
\bibliography{Mbibliography}

\end{document}